\documentclass[fleqn,usenatbib]{mnras}

\usepackage{graphicx}	% Including figure files
\usepackage{amsmath}	% Advanced maths commands
\usepackage{amssymb}	% Extra maths symbols
\usepackage{multicol}        % Multi-column entries in tables
\usepackage{bm}		% Bold maths symbols, including upright Greek
\usepackage{pdflscape}	% Landscape pages
\usepackage{xcolor}
\usepackage{soul}

\usepackage[T1]{fontenc}
\usepackage{ae,aecompl}

\usepackage{newtxtext,newtxmath}
\title[VLBI-detected source counts]{On the source counts of VLBI-detected radio sources and the prospects of all-sky surveys with current and next generation instruments}

\author[S. Rezaei et al.]{
S. Rezaei,$^{1,2}$\thanks{E-mail: rezaei@strw.leidenuniv.nl}
J. P. McKean$^{3,4}$,
A. T. Deller$^{5}$ and J. F. Radcliffe$^{6,7}$
\\
$^{1}$Leiden Observatory, Leiden University, Box 9513, 2300 RA Leiden, The Netherlands\\
$^{2}$Leiden Institute of Advanced Computer Science (LIACS), Leiden University, P.O. Box 9512, 2300 RA Leiden, The Netherlands\\
$^{3}$Kapteyn Astronomical Institute, University of Groningen, Postbus 800, NL-9700 AV Groningen, The Netherlands\\
$^{4}$ASTRON, Institute for Radio Astronomy, Oude Hoogeveensedijk 4, 7991 PD Dwingeloo, The Netherlands\\
$^{5}$Centre for Astrophysics and Supercomputing, Swinburne University of Technology, Mail H30, PO Box 218, Hawthorn, VIC 3122, Australia \\
$^{6}$Department of Physics, University of Pretoria, Lynnwood Road, Hatfield, Pretoria, 0083, South Africa\\
$^{7}$Jodrell Bank Centre for Astrophysics, University of Manchester, Oxford Road, Manchester M13 9PL, UK
}

\date{Accepted XXX. Received YYY; in original form ZZZ}

\pubyear{2023}

\begin{document}
\label{firstpage}
\pagerange{\pageref{firstpage}--\pageref{lastpage}}
\maketitle

% Abstract of the paper
\begin{abstract}
We present an analysis of the detection fraction and the number counts of radio sources imaged with Very Long Baseline Interferometry (VLBI) at 1.4 GHz as part of the mJIVE--20 survey. From a sample of 24\,903 radio sources identified by FIRST, 4\,965 are detected on VLBI-scales, giving an overall detection fraction of $19.9\pm2.9$~per cent. However, we find that the detection fraction falls from around 50 per cent at a peak surface brightness of 80~mJy~beam$^{-1}$ in FIRST to around 8 per cent at the detection limit, which is likely dominated by the surface brightness sensitivity of the VLBI observations, with some contribution from a change in the radio source population. We also find that compactness at arcsec-scales is the dominant factor in determining whether a radio source is detected with VLBI, and that the median size of the VLBI-detected radio sources is 7.7 mas. After correcting for the survey completeness and effective sky area, we determine the slope of the differential number counts of VLBI-detected radio sources with  flux densities $S_{\rm 1.4~GHz} > 1$~mJy to be $\eta_{\rm VLBI} = -1.74\pm 0.02$, which is shallower than in the cases of the FIRST parent population ($\eta_{\rm FIRST} = -1.77\pm 0.02$) and for compact radio sources selected at higher frequencies ($\eta_{\rm JBF} = -2.06\pm 0.02$). From this, we find that all-sky ($3\pi$~sr) surveys with the EVN and the VLBA have the potential to detect $(7.2\pm0.9)\times10^{5}$ radio sources at mas-resolution, and that the density of compact radio sources is sufficient (5.3~deg$^{-2}$) for in-beam phase referencing with multiple sources (3.9 per primary beam) in the case of a hypothetical SKA-VLBI array.
%next generation instruments, such as SKA-VLBI and the ngVLA, can potentially detected a factor ?? more.
%In this study we analysed the mJy radio sources population by Very Large Baseline Array (VLBA). Our data samples are chosen from mJy Imaging VLBA Exploration at 20 cm (mJIVE-20) which is based on Faint Images of the Radio Sky at Twenty cm (FIRST) survey. As only a small portion of the existing sources have been detected by mJIVE-20, this study focuses on the differences between detected and non detected radio populations of this survey. We compared the density of detected objects in mJIVE-20 and corresponding FIRST parenting samples by calculating their number count. The results are quite interesting and informative for making choices for future larger surveys.
\end{abstract}

% Select between one and six entries from the list of approved keywords.
% Don't make up new ones.

%%%%%%%%%%%%%%%%%%%%%%%%%%%%%%%%%%%%%%%%%%%%%%%%%%

%%%%%%%%%%%%%%%%% BODY OF PAPER %%%%%%%%%%%%%%%%%%

% The MNRAS class isn't designed to include a table of contents, but for this document one is useful.
% I therefore have to do some kludging to make it work without masses of blank space.

\section{Introduction}

Very Long Baseline Interferometry (VLBI; \citealt{Broten1967}) is an observational technique where the signals from individual radio telescopes are combined coherently to produce a synthesised unfilled aperture. As the radio telescopes can be widely separated across the Earth (and even located in space), this technique currently provides the highest angular resolution imaging in astronomy (with a synthesized beam size of typically 1 to 10 mas at cm-wavelengths, and reaching 0.025 mas at sub-mm wavelengths; \citealt{EHT2019b}). VLBI allows a wide range of science goals to be realised, for example, detecting super-luminal motion in radio jets \citep[e.g.,][]{Cohen1977}, imaging gravitational lenses \citep[e.g.,][]{Porcas1979}, mapping the accretion disks of supermassive black holes \citep[e.g.,][]{Miyoshi1995}, tracing the expansion of stellar explosions \citep[e.g.,][]{OBrien2006}, imaging outflows of atomic hydrogen from supermassive black holes \citep[e.g.,][]{Morganti2013}, localising Fast Radio Bursts \citep[e.g.,][]{Marcote2020}, and making the first images of the shadow of a black hole \citep[e.g.,][]{EHT2019a}.   

It is over 50 years since the first fringes were produced between two unconnected antennas separated by over 3000~km \citep{Broten1967}, yet the unique applications of VLBI at cm-wavelengths have been restricted to studying only $\sim25\,000$ radio sources with very high brightness temperatures ($>10^5$~K) and over a very small fraction of the observable sky ($\sim 200$~deg$^2$; e.g., \citealt{Deller2014,Herrera_Ruiz2017,Petrov2021}). This is significantly lower than the up to 5 million radio sources that have been catalogued from recent all-sky surveys at arcsec-resolution (e.g., \citealt{Becker1995,Condon1998,Intema2017,Shimwell2019,Lacy2020}). This is because the effective field-of-view of a VLBI experiment is typically only a few tens of arcsec in diameter, which is due to historical computational limitations and data-rates that required significant averaging of the visibility data. This means that large numbers of sources can only be observed through many short observations. Coupled with the sparseness of the telescopes forming the synthesised unfilled aperture, this results in only the brightest sources being detectable, that is, those that are sufficiently compact to have a measurable correlated flux on the available baselines of the array.

This limitation of VLBI means that radio sources at relatively high flux densities ($>1$~mJy) tend to make up the target population at cm-wavelengths, which is mainly dominated by radio-loud Active Galactic Nuclei (AGN), or so-called jetted AGN \citep{Padovani2017}. At lower flux densities  ($<1$~mJy) a higher proportion of radio sources have emission associated with star-formation processes or weak jets (e.g.,\citealt{Condon2012,Prandoni2018}). Studying such objects at high angular resolution requires much deeper observations, which further limits the number of objects that can be studied \citep{Garrett2001}.

However, due to advances in computing and the development of new techniques for correlating the signals from the different radio telescopes, it is now possible to form multiple phase centres for a given observation \citep{Deller2011}. Using multiple phase centres allows the observer to image small areas around known sources within the same observation. It works by shifting the phase centre to the location of known sources and averaging the obtained visibilities to obtain manageable-sized data sets. For example, the GOODS-N \citep{Radcliffe2018} and COSMOS fields had 699 and 750 phase centres per observation. However, these are somewhat special cases and routine wide-field VLBI experiments are currently limited to around 100 phase centres from a typical observation. Nevertheless, this technique has revolutionised our view of the mas-scale radio Universe, both through shallow wide-field surveys, like the mJy Imaging VLBI Exploration at 20 cm (mJIVE--20; \citealt{Deller2014}) survey and through deep narrow-field observations of well-studied deep fields \citep[e.g.,][]{Middelberg2013,Herrera_Ruiz2017,Radcliffe2018}. 

The results from these studies have demonstrated a proof-of-concept for all-sky VLBI surveys with current instruments, like the European VLBI Network (EVN) or the Very Long Baseline Array (VLBA). Also, in the future, the sensitivity of VLBI arrays are expected to improve dramatically with the construction of the Square Kilometre Array (SKA-VLBI) and the next generation Very Large Array (ngVLA). These instruments will lower the current sensitivity limits, and given their wide fields-of-view and wide frequency bandwidths, they can provide a significant increase in the numbers of sources detected on arcsec- to mas-scales. However, this will require determining the number of detectable sources on VLBI-scales as a function of flux density, so that survey strategies can be properly developed.

In this paper, we determine the source counts of VLBI-detected radio sources from the wide-area mJIVE--20 survey \citep{Deller2014}. With these data, we estimate the number of radio sources that can be detected by the next generation of wide-field VLBI surveys with the VLBA and the EVN in the short term ($<10$~yr) and with a hypothetical SKA-VLBI array in the longer term. This paper is arranged as follows. In Section~\ref{mJIVE20}, we present an overview of the final mJIVE--20 survey catalogue that we use for our study. We determine the source counts of VLBI-detected radio sources in Section \ref{NC}, which includes calculating the mJIVE--20 survey completeness from simulations and determining the effective area of the survey. 
%In Section \ref{optical}, we investigate the properties of the VLBI detected/non-detected sub-samples, with respect to their optical emission. 
Using these data and the (expected) capabilities of current and next generation radio interferometers, we also calculate the number of likely detectable radio sources on VLBI-scales for given survey strategies in Section~\ref{discussion}. Finally, in Section~\ref{conclusions}, we present our conclusions. 

\section{The mJIVE--20 survey final catalogue}
\label{mJIVE20}

In this section, we provide an overview of the mJIVE--20 survey, and discuss the properties and detection rate of the final catalogue that we use to determine the source counts of VLBI-detected radio sources.

\subsection{Overview of the mJIVE--20 survey}

The mJIVE--20 survey was carried out with the VLBA at 1.4 GHz with dual polarization and a bandwidth of 64 MHz (recording rate of 512 Mbit\,s$^{-1}$). The main aim of the project was to better understand the radio source population at mas resolution by exploiting recent developments in wide-field multi-phase centre correlation techniques \citep{Deller2011}. The survey strategy and initial results were reported by \citet{Deller2014}; here we summarise the main properties of the survey and discuss the final catalogue. 

\begin{figure}
 \includegraphics[width=\columnwidth]{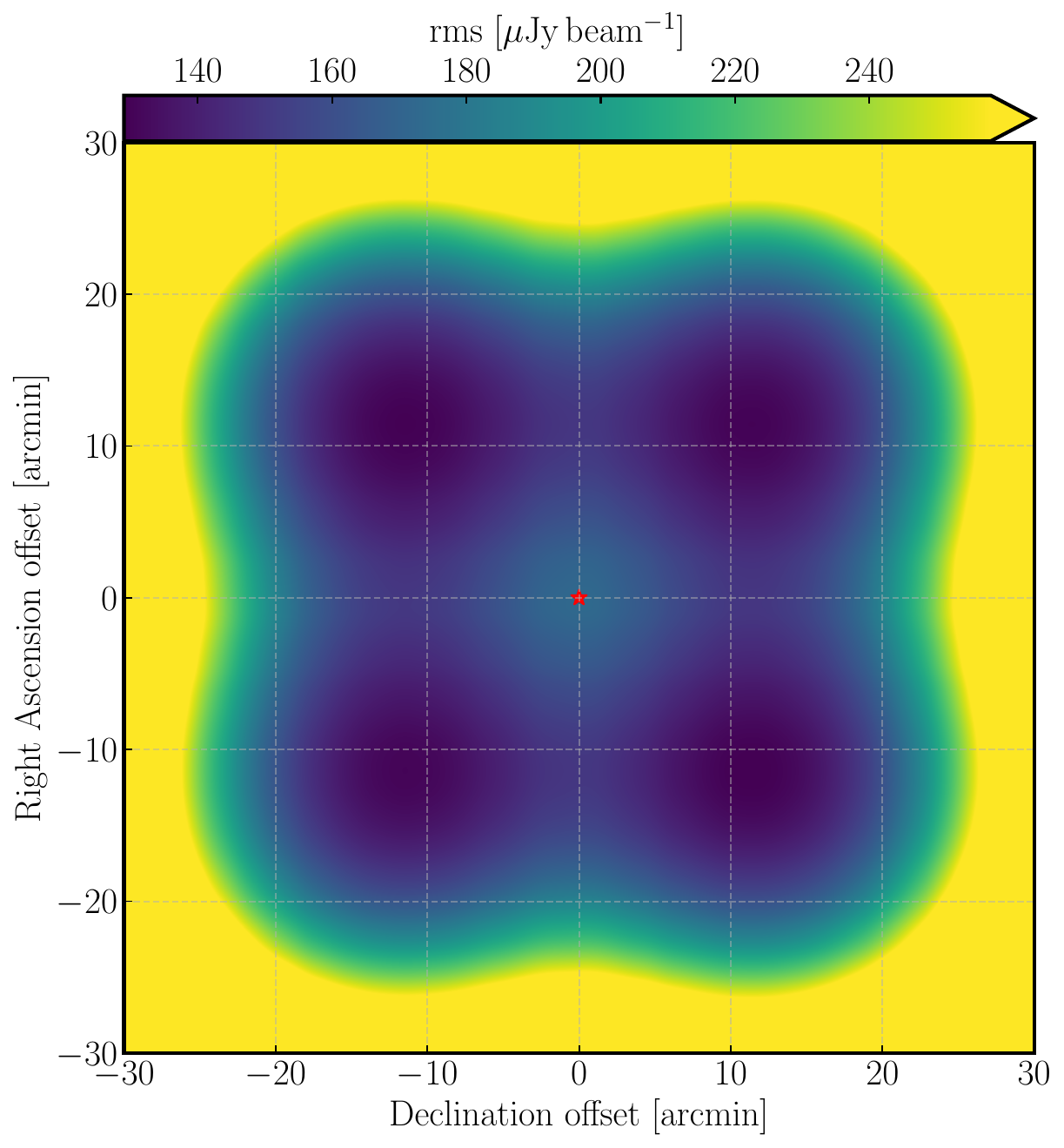}
 \caption{An example rms map from simulating four separate pointings of the VLBA 25-m antennas at 1.4 GHz. The VLBA calibrator source is denoted by the (red) star in the centre of the image, such that it is always in the primary beam of each antenna for each pointing position. A sky area of about $1\times1$~deg$^2$ is surveyed with this pointing strategy, for each observation of a VLBA calibrator.}
 \label{fig:area}
\end{figure}

Each mJIVE--20 survey observation consisted of four individual pointings of 2 mins duration around a known VLBA calibrator (so that in-beam calibration could be used; see Fig.~\ref{fig:area}) at 7 different hour angles to improve the {\it uv}-coverage. This resulted in a typical synthesised beam-size of $16\times6$~mas for target fields at Declinations between $-7$ and 60 deg. The average rms noise is around 150~$\mu$Jy~beam$^{-1}$ at each pointing centre. In total, there were 410 separate observations that targeted 24\,903 radio sources from the Faint Images of the Radio Sky at Twenty-cm (FIRST) survey \citep{Becker1995}, making the mJIVE--20 survey the largest targeted sample of radio sources with VLBI to date at L band. From these observations, 4\,965 sources were detected above a threshold of 6.75$\sigma$, where $\sigma$ is the local rms noise of the imaging data. For our analysis, we use the final source catalogue from 2014 March 31, which was obtained using the {\sc blobcat} source detection software \citep{Hales2012}.

\subsection{Detection rate as a function of radio source surface brightness, compactness and size}

We now investigate the detectability of radio sources on VLBI-scales, based on their observed properties at lower angular resolution. Note that this comparison is not intended to be a thorough study of the astrophysical characteristics of the radio source population, but instead aims to compare the properties between the parent sample catalogue obtained with the VLA and whether a detection would be made with a snapshot observation with the VLBA.

\begin{figure}
\includegraphics[width=\columnwidth]{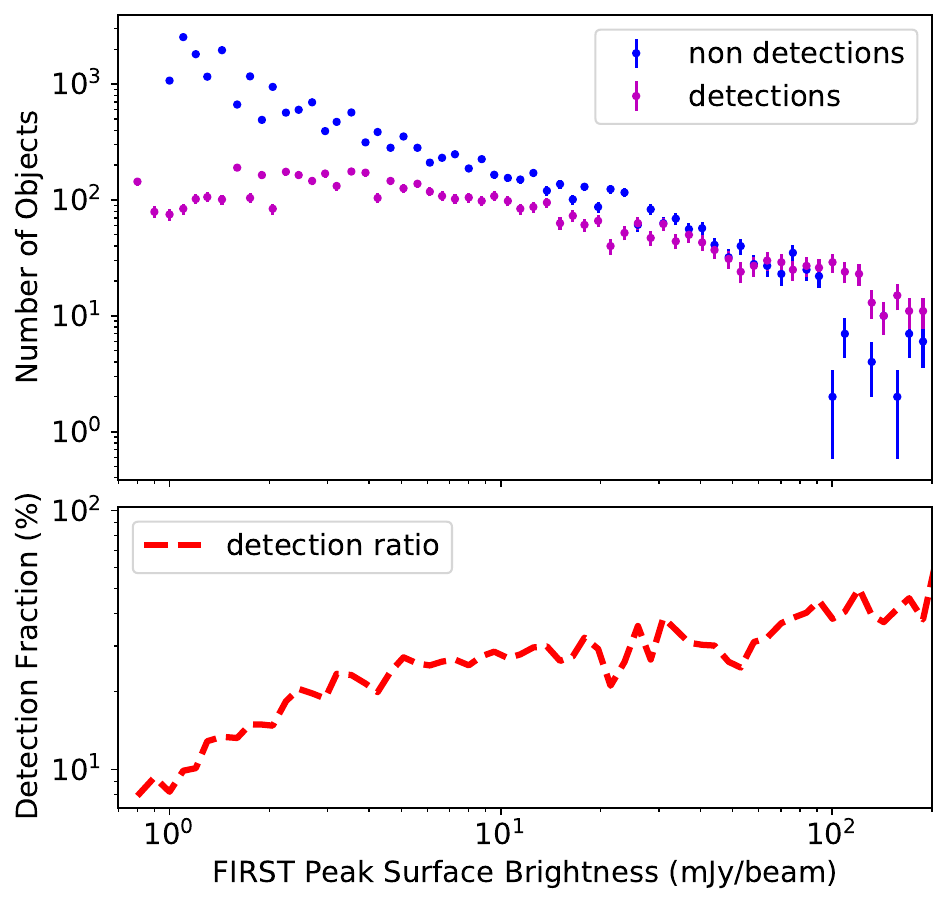}
\caption{Upper panel: The number of objects detected (purple) and not-detected (blue) at VLBI-scales during the mJIVE--20 survey, as a function of peak surface brightness measured from FIRST (5.5 arcsec beam size). Above a peak brightness of around 90 mJy~beam$^{-1}$, the number of detections is greater than the number of non-detections. Lower panel: The detection fraction at VLBI-scales during the mJIVE--20 survey, as a function of peak surface brightness measured from FIRST. The detection rate changes from around 50 per cent for the highest surface brightness sources within FIRST, to below 10 per cent at the detection threshold. This is due to the compactness of the radio sources changing as a function of flux density.}
\label{fig:mjivedetectionratio}
\end{figure}

From a straight comparison of the number of sources observed and detected, we see that the mJIVE--20 survey has an overall detection rate of $19.9\pm2.9$~per cent, where the uncertainty is calculated from Poisson statistics. However, to investigate the detection rate of radio sources on VLBI-scales further, we show in Fig.~\ref{fig:mjivedetectionratio} the number of radio sources detected and not detected, and the detection fraction for the mJIVE--20 survey, as a function of the FIRST peak surface brightness. We see that above a peak surface brightness of 80 mJy~beam$^{-1}$ (for a 5.5 arcsec beam size), the majority of radio sources detected in FIRST also have a VLBI counterpart in the mJIVE--20 survey. Below this surface brightness limit, the detection fraction steadily drops from 50 to around 31 per cent at 5 mJy~beam$^{-1}$, before steeply falling to 8 per cent at the FIRST detection limit (5$\sigma$) of 0.8 mJy~beam$^{-1}$. 

This change in the detection fraction as a function of peak surface brightness in the parent sample could result from a number of factors. First, it could be due to a change in the intrinsic compactness of radio sources, where the brightest objects are relativistically beamed towards the observer, that is, a selection effect. This would explain why the majority of radio sources at the highest surface brightness are also detected on VLBI-scales. Also, even for those cases that are not strongly beamed, the depth of the mJIVE--20 survey data is sufficient to detect a weak radio core in the brightest radio sources in FIRST. This is consistent with the results of \citet{Herrera_Ruiz2017}, who detected with the VLBA a large number of low flux density radio sources ($\leq$1 mJy) that were previously identified using the VLA. Moreover, for those observations, the VLBA recovered between 60 and 80 per cent of the VLA flux density. This suggests that a large portion of sources at and below the mJIVE--20 survey detection limit may still be dominated by AGN activity.

Second, it could be that those sources with a lower surface brightness do have radio emission on VLBI-scales, but the mJIVE--20 survey observations were not deep enough. This would explain the sharp drop in the detection rate towards a lower surface brightness, that is, this is an observational effect. This conclusion is also consistent with extremely deep EVN observations of the GOODS-N field (rms 9~$\mu$Jy~beam$^{-1}$; \citealt{Radcliffe2018}) when an \textit{e}-MERLIN/VLA parent sample is used \citep{Muxlow2020}, which finds a similar detection rate of $25\pm6$ per cent (24 out of 94 objects) to the mJIVE--20 survey. Even though the radio source population at this depth is expected to be dominated by star-forming galaxies, as opposed to those harbouring an AGN, it is interesting that providing deep enough observations are carried out, a high fraction of VLBI detections is still found. Therefore, the steep change in the detection fraction towards lower surface brightness for the mJIVE--20 sample is likely a combination of the observing depth and a change in the properties of the underlying source population. %This is discussed further in Section~\ref{discussion}.

\begin{figure}
 \includegraphics[width=\columnwidth]{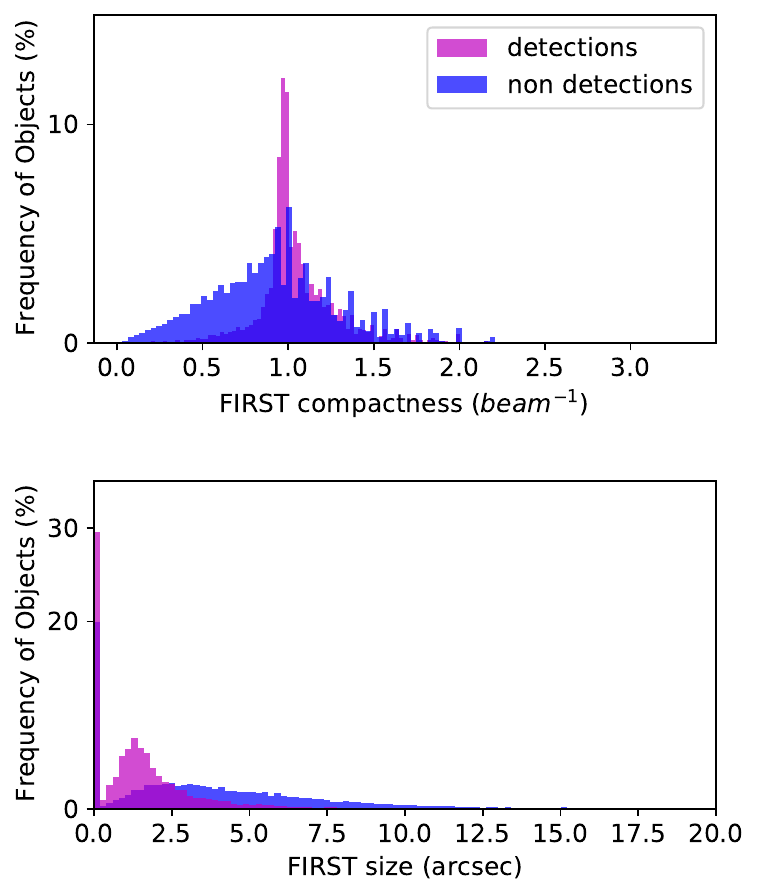}
 \caption{The compactness ($F_{\rm comp}$; upper panel) and major-axis size ($F_{\rm size}$; lower panel) in the FIRST survey for the detection (purple) and non-detection (blue) samples in the mJIVE--20 survey.}
 \label{fig:FITSThist}
\end{figure}

\begin{figure}
\centering
 \includegraphics[width=\columnwidth]{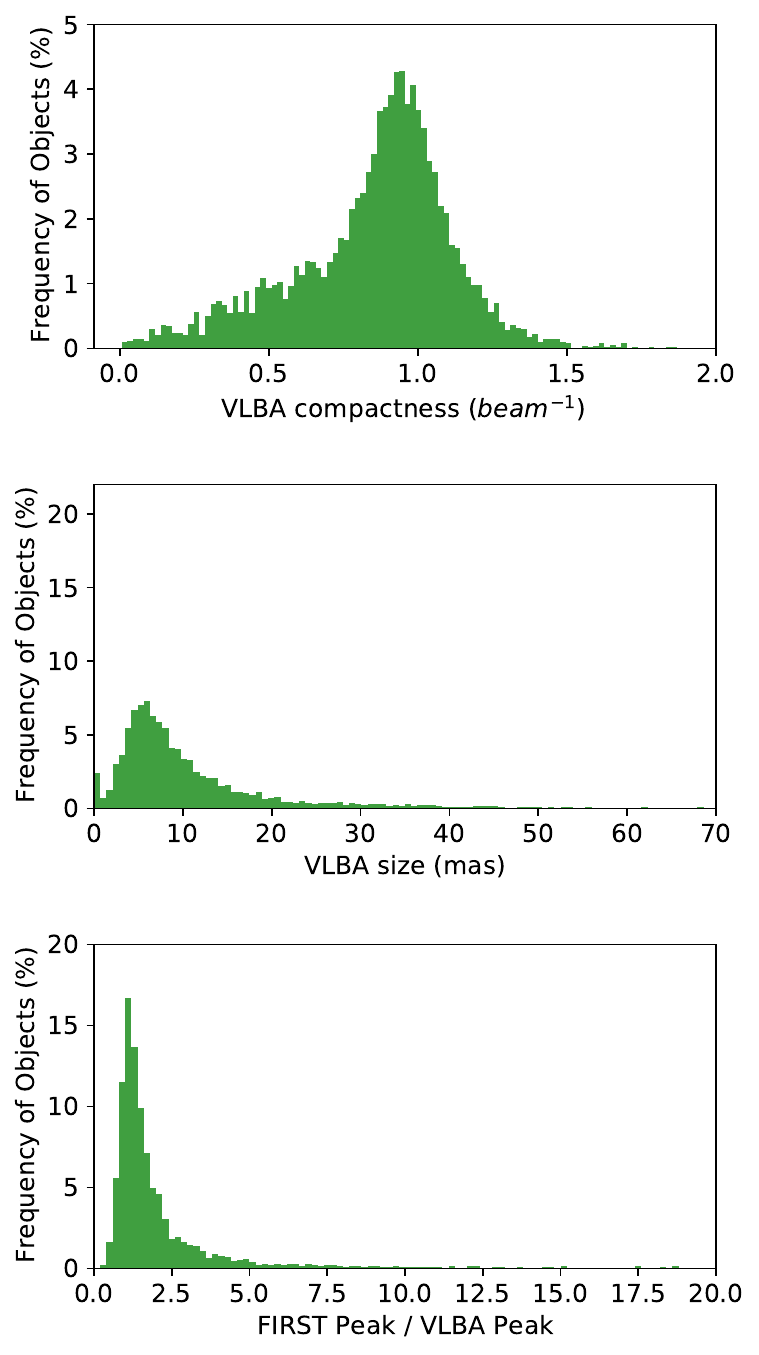}
 \caption{The compactness ($V_{\rm comp}$; upper panel) and major-axis size ($V_{\rm size}$; middle panel) for objects detected in the mJIVE--20 survey. Also shown is the ratio of the FIRST and mJIVE--20 survey peak surface brightness ($P_{F/V}$; lower panel).}
 \label{fig:VLBIhist}
\end{figure}

To further investigate the latter, we have looked at several properties of the source population to determine their effect on the mJIVE--20 survey detection fraction. In particular, we have analysed the ratio of VLBI peak surface brightness to VLBI integrated flux density (VLBA compactness), the ratio of FIRST peak surface brightness to FIRST integrated flux density (VLA compactness), and the VLBI and FIRST deconvolved major axis (taken from the mJIVE--20 survey catalog obtained using {\sc blobcat}), which are defined as $V_{\rm comp}$, $F_{\rm comp}$, $V_{\rm size}$ and $F_{\rm size}$, respectively. Distributions for the VLA compactness and size are presented in Fig.~\ref{fig:FITSThist}. The same distributions for the VLBA properties, and also the ratio of peak VLBI surface brightness to the peak FIRST surface brightness, which we refer to as $P_{F/V}$ (radio source compactness), are presented in Fig.~\ref{fig:VLBIhist}.

As expected, 
%we see from Fig.~\ref{fig:FITSThist} that 
those objects with a higher compactness and a smaller size on VLA-scales are more likely to also have a detection on VLBI-scales. For the VLA compactness, the mean (with standard deviation) and median of the distributions for the VLBI detections and non-detections are 1.05 ($\sigma = 0.24$) and 1.00 beam$^{-1}$, and 0.92 ($\sigma = 0.38$) and 0.92 beam$^{-1}$, respectively. In the case of the size, the mean (with standard deviation) and median of the distributions for the VLBI detections and non-detections are  1.60 ($\sigma =2.01$) and 1.24 arcsec, and 4.10 ($\sigma =4.10$) and 3.27 arcsec, respectively. Also, we find that those objects detected on VLBI-scales by the mJIVE--20 survey tend to have a higher compactness and smaller intrinsic size; the mean (with standard deviation) and median for the compactness and size on VLBI-scales is 0.85 ($\sigma = 0.27$) and 0.9 beam$^{-1}$, and 11.5 ($\sigma = 13.0$) and 7.7 mas. Finally, we find that $P_{F/V}$ has a mean of 3.7, with a standard deviation of 13.4, and a median of 1.4.

\begin{figure}
 \includegraphics[width=\columnwidth]{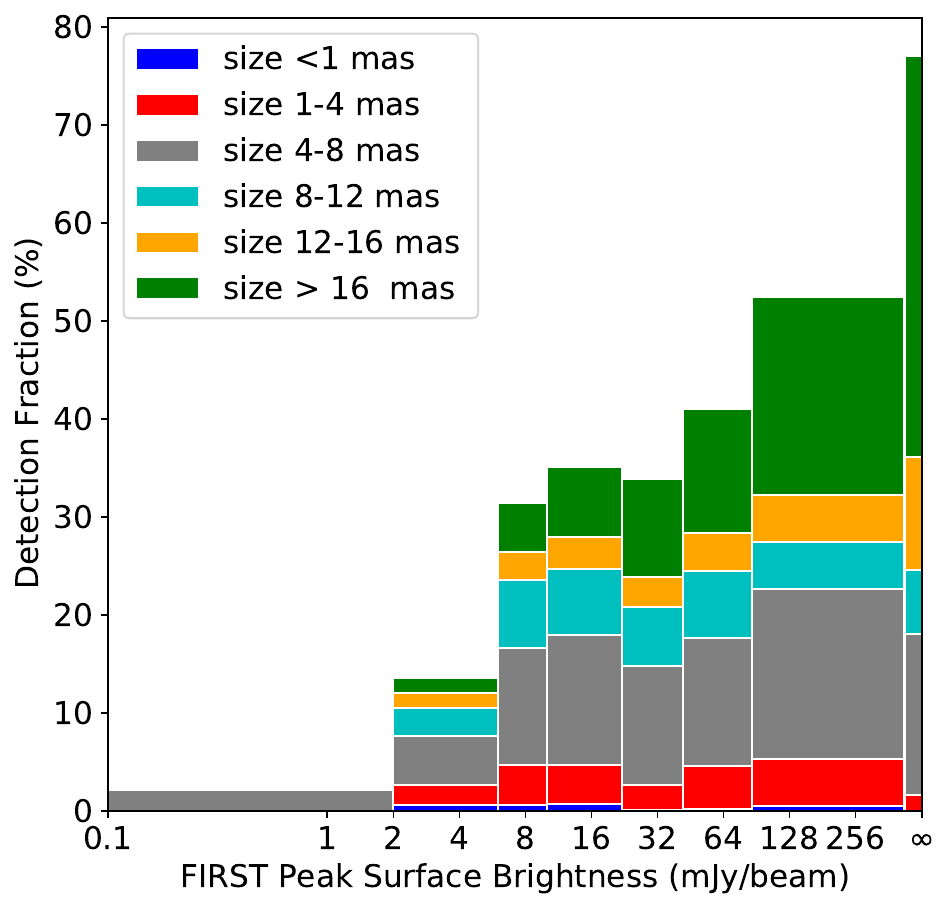}
 \caption{The detection fraction for the mJIVE--20 survey, binned by peak surface brightness in FIRST. The different colours show the detection fraction in bins of  $V_{\rm size}$ (deconvolved major axis of a single 2-dimensional elliptical Gaussian fit to the imaging data).}
 \label{fig:VLBIsizebmaj}
\end{figure}

Although the mean values of the above parameters tend to be consistent for the samples of detections and non-detections on VLBI-scales, given the large scatter for each parameter, the median values point toward compact sources being more likely to be detected. Therefore, we now investigate the distribution of sources detected, as a function of their VLBI size, $V_{\rm size}$. In Fig.~\ref{fig:VLBIsizebmaj}, we show the mJIVE--20 survey detection fraction as a function of FIRST peak surface brightness, where each bin is also divided into bins of $V_{\rm size}$. We consider all objects with a major axis larger than 16 mas as extended on VLBI-scales, since this is equivalent to the average beam size of the mJIVE--20 survey. Due to the selection bias in the low surface brightness regime ($<2$~mJy~beam$^{-1}$ in FIRST), only objects with a $V_{\rm size}$ of $<8$~mas have been detected. Overall, extremely compact objects ($V_{\rm size} <1$~mas) have the least contribution to the detection fraction compared to the more extended objects, implying that almost all radio sources are partially-resolved with the VLBA, that is, they have some level of VLBI structure. Another point from Fig.~\ref{fig:VLBIsizebmaj} is that for a surface brightness $>8$~mJy~beam$^{-1}$, the contribution of objects with $1< V_{\rm size} <16$~mas is almost constant. However, the contribution of those objects with $V_{\rm size} > 16$~mas increases. This is likely due to the signal-to-noise ratio of the brightest radio sources also being high, and therefore, we are more sensitive to any possible extended emission. Similarly, extended objects make up a smaller fraction of the lowest surface brightness part of the distribution because here we are mainly sensitive to the most compact emission from an object.

Overall, we find that VLBI detections are more likely from the most compact radio sources observed at arcsec-scales, and that the emission observed on VLBI-scales for around 65 to 80 per cent of the objects has a size in the range of 1 to 16 mas.

\section{VLBI-detected radio source counts}
\label{NC}

In this section, we determine the source counts of radio sources detected during the wide-field mJIVE--20 survey. Our aim is to estimate the number of VLBI-detected radio sources as a function of integrated flux density, so that we can make robust estimates for the likely source counts from future surveys with current (EVN and VLBA) and next generation (SKA-VLBI and ngVLA) VLBI arrays.

We first calculate the sky-area observed during the mJIVE--20 survey, which due to the primary beam attenuation of the VLBA antennas is also a function of source surface brightness. Next we determine the completeness of the catalogue by making simulations of mock mJIVE--20 survey data and running the source detection procedure. With the sky area and an estimate of the completeness in hand, we then determine the normalized Euclidean and differential source counts for VLBI-detected radio sources. Here, we refer to the entire sample of 24\,903 radio sources observed during the mJIVE--20 survey as the (FIRST) parent population, and the 4\,965 radio sources detected with the VLBA as the mJIVE--20 population.

\subsection{The mJIVE--20 survey sky area}

Calculating the radio source density on the sky requires some knowledge of the sky area being investigated. As can be seen in Fig.~\ref{fig:area}, each mJIVE--20 survey observation consisted of four different pointing positions around a central calibrator source \citep{Deller2014}. This resulted in the effective rms noise across the field being non-uniform, due to the primary beam attenuation of the individual antennas and the combination of data from different individual pointings and repeated observations. Also, two different pointing configurations were employed during the mJIVE--20 survey. The first used an offset from the central calibrator of $\pm12$ arcmin in RA and Dec, to include all parent population radio sources within 20 arcmin of each pointing centre, and the second used offsets of $\pm9.6$ arcmin in RA and Dec, and included all parent population radio sources within 17 arcmin of each pointing centre. A total of 306 unique calibrators were observed, with 252 observations using the set-up with $\pm12$ arcmin offsets, and 54 observations using the set-up with $\pm9.6$ arcmin offsets from the calibrator.

To measure the sky area, we use a Monte Carlo integration by stone throwing technique, which is based on the mean value theorem designed for numerical integration using random numbers. In this method, the sky area is calculated by the average number of hits in different iterations. The error in measuring the sky area with this approach goes as ${1}/{\sqrt{N}}$, where $N$ is the number of thrown stones. We first define a box around the intersected primary beam footprints (circles), shown for example in Fig. \ref{fig:area}, and measure the area of that box. Then, we throw 1 million points into the box to estimate the size. The ratio of the number of points that are inside any of the circles to the total number of points in the box gives an estimation of the area inside the circles. Considering the area within the individual observation footprints, and the different pointing configurations, the total sky area of the mJIVE--20 survey is found to be 237.95~deg$^2$. 

\begin{figure}
 \includegraphics[width=\columnwidth]{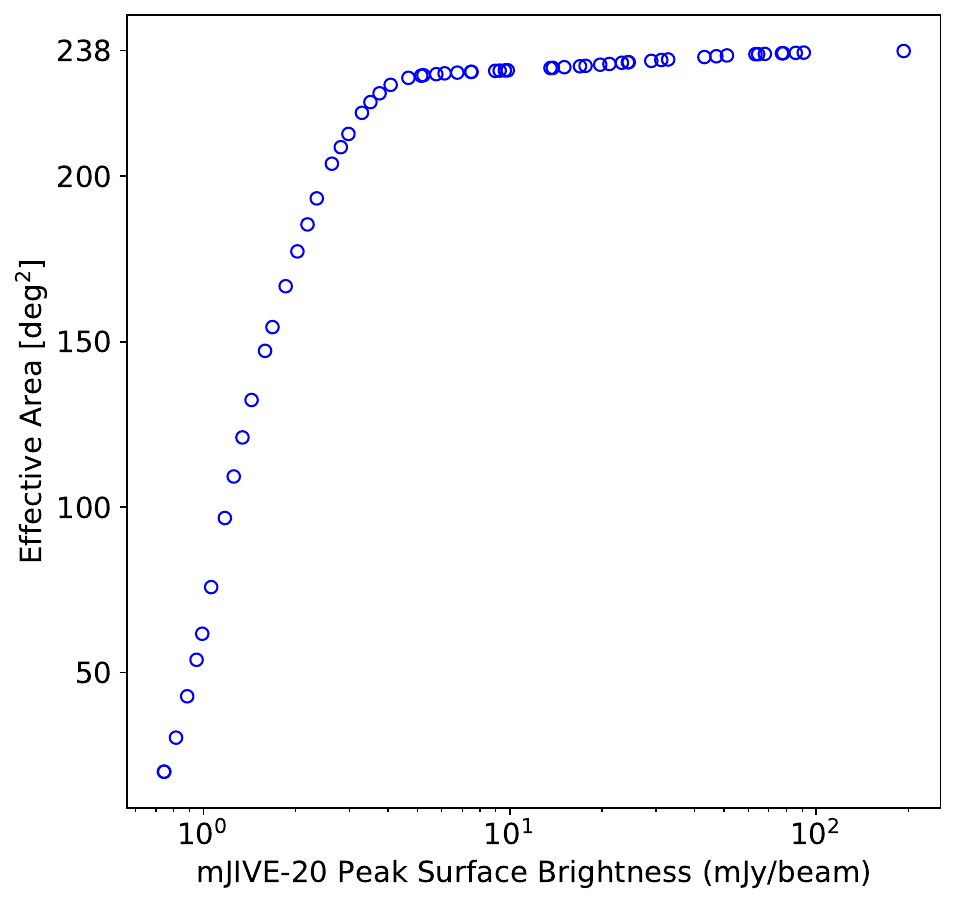}
 \caption{The effective area of the mJIVE--20 survey, as a function of minimum detectable radio source surface brightness (for a detection threshold of $6.75\sigma$). The total observed area in the mJIVE--20 survey is 237.95~deg$^{2}$, but changes as a function of source brightness due to the primary beam attenuation and the combination of the visibility data from different pointings and repeated observations.}
 \label{fig:mJIVE_effectivearea}
\end{figure}

In Fig.~\ref{fig:mJIVE_effectivearea}, we show how the effective area changes as a function of the minimum peak surface brightness of an mJIVE--20 population radio source (assuming a local signal-to-noise ratio of 6.75). We see that at a surface brightness of around 4 mJy~beam$^{-1}$, there is a knee in the effective sky area (corresponding to 227~deg$^2$; 95 per cent of the total sky observed), which is roughly equivalent to 25 times the typical rms map noise of 150~$\mu$Jy~beam$^{-1}$ at each pointing centre.

\subsection{The mJIVE--20 survey completeness}

\begin{figure}
 \includegraphics[width=\columnwidth]{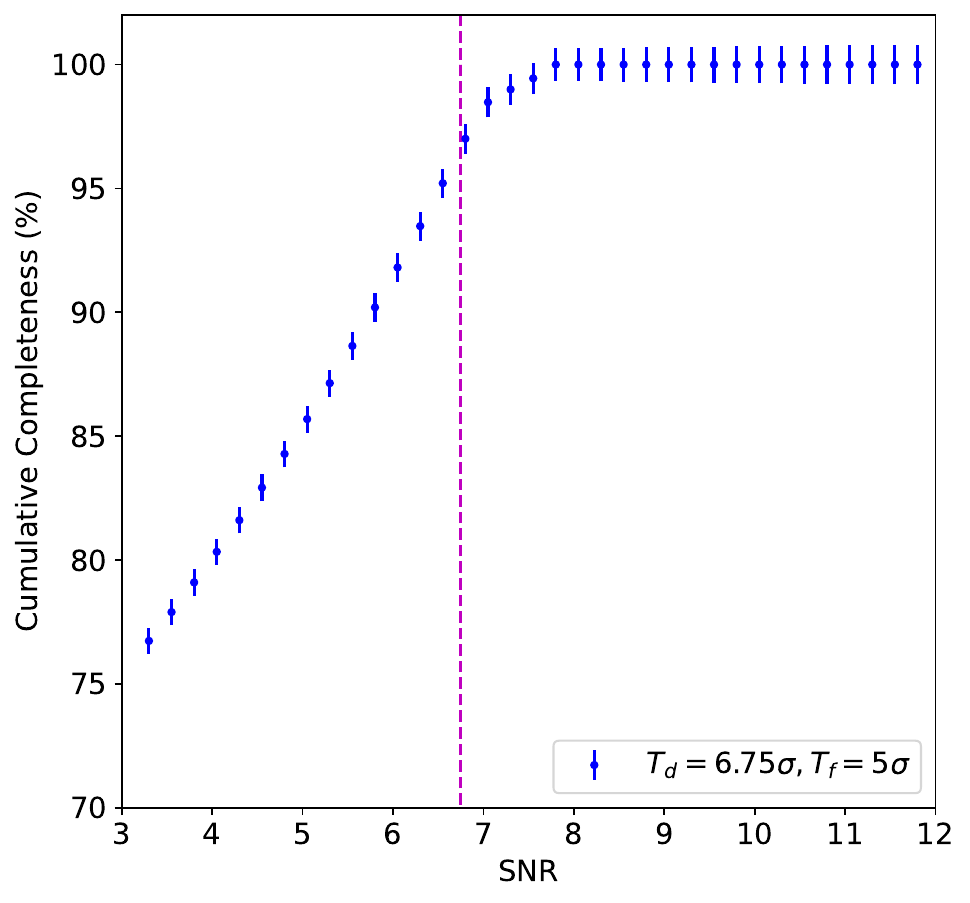}
 \caption{The cumulative completeness for a set of simulated mock VLBA observations that have the same characteristics as the mJIVE--20 survey data. These have been analyzed using the {\sc blobcat} object detection algorithm, with the same parameters for the detection thresholds ($T_d$ and $T_f$; see text for details) used for the mJIVE--20 survey.}
 \label{fig:blobcat_mjivesetting}
\end{figure}

In order to calculate the source counts of the mJIVE--20 population to the lowest flux densities, we also have to consider the completeness of the catalogue, that is, estimate how many sources may have been missed due to observational effects. To do this, we have made mock visibility data sets, which then go through the same imaging and source detection process as the mJIVE--20 survey data. By comparing the input and output mock catalogues, we determine the completeness as a function of flux density.

The simulated data set needed to have similar properties to the radio sources in the mJIVE--20 population catalogue. Therefore, we implemented a Monte Carlo approach to define the simulated radio source size (either delta or 2-dimensional elliptical Gaussian function), peak surface brightness, position angle, and a random $x, y$ position with respect to the phase centre, given the actual distribution of these parameters in the mJIVE--20 catalogue (see \citet{Rezaei2021} for a detailed overview of generating realistic mock sources for mJIVE--20 survey). The generated sources were then converted to mock visibilities using the Common Astronomy Software Applications ({\sc casa}; \citealt{McMullin2007}) package. The simulation tool in {\sc casa} allows as an input an existing interferometric visibility data set that can be used to obtain the position of the antennas and other observational settings (such as frequency and time sampling, total integration time). Therefore, we used the actual mJIVE--20 survey interferometric visibility data sets when converting the simulated model radio sources into mock visibility data. The final step in the process used the tclean task within {\sc casa} to produce de-convolved clean images. As with the mJIVE--20 survey data, the images were centred on the position of the surface brightness peak, had a size of $1024\times1024$ pixels and a pixel-scale of 0.75 mas~pixel$^{-1}$. The dirty images were cleaned for a maximum of 1000 iterations or until a threshold of 0.2 mJy~beam$^{-1}$ was reached, while masking was applied to the centre of the image, with a radius of 25 pixels.

To identify and measure the flux density of the sources in our mock imaging data, we have used {\sc blobcat} \citep{Hales2012}, the same post-processing object detection package used as part of the mJIVE--20 survey. {\sc blobcat} searches for islands of pixels that could represent a possible object considering the signal-to-noise ratio of a given pixel. This is accomplished using the surface brightness distribution of the 2-dimensional imaging data and the background rms noise as the input parameters. It requires the user to set a threshold for the minimum detection signal-to-noise ratio (known as $T_d$) as well as a cutoff ratio (known as $T_f$) for flooding the islands. Based on simulations carried out by \citet{Deller2014}, $T_d$ and $T_f$ were set to $6.5\sigma$ and $5\sigma$, respectively, during the mJIVE--20 survey. However, only those sources with a peak surface brightness above $6.75\sigma$ were added to the final mJIVE--20 survey catalogue, so we adopt that value for $T_d$ in our simulations. In Fig.~\ref{fig:blobcat_mjivesetting}, we show the cumulative completeness of the mJIVE--20 survey as a function of signal-to-noise ratio from 13\,500 simulated radio sources that have a signal-to-noise ratio between 3 and 15. The magenta dashed line shows the $6.75\sigma$ detection threshold of the mJIVE--20 survey. We find from our simulations that {\sc blobcat} is 97 per cent complete at a threshold of $6.75\sigma$ and reaches full completeness at a threshold of $7.8\sigma$ for this data set. 

From Fig.~\ref{fig:blobcat_mjivesetting} we are able to calculate the completeness of the mJIVE--20 survey, given the observational set-up (signal-to-noise ratio, {\it uv}-coverage, image de-convolution) and the ability of the object detection algorithm to identify a representative sample of realistic radio sources. This allows us to define a completeness correction factor for each flux density bin, where those bins below 100 per cent completeness are divided by the completeness given in Fig.~\ref{fig:blobcat_mjivesetting}; for this, we assume that the radio sources are mostly unresolved, which is consistent with the results presented in the previous section.

\subsection{A note on the resolution bias of the parent and mJIVE--20 population samples}

As the mJIVE--20 survey is not a blind search on VLBI-scales, but is instead a targeted search of known FIRST radio sources, the completeness of that parent population can also affect the final source counts of VLBI-detected radio sources. In fact, it is known that below a flux density of around 2 mJy, the completeness of the FIRST survey becomes progressively worse \citep{White1997}, and the observed source counts dramatically drop-off from what is predicted from models of radio source evolution and observations of deep fields at the same frequency (e.g., \citealt{Prandoni2018}). This results in the source counts of FIRST being lower by a factor of between 0.9 and 0.4 towards the faint end of the flux density distribution. However, as this effect is thought to be due to faint and extended radio sources being resolved out by the 5.5 arcsec synthesised beam of the VLA (B-configuration; see \citealt{White1997} for further discussion), it is almost certainly the case that such objects would also be resolved out at VLBI-scales (in the absence of significant variability). Therefore, we do not correct for the resolution bias of the FIRST survey.

However, there will be a resolution bias associated with the mJIVE--20 survey data, as the detectability and measured flux density of a given object will be dependent on the structure (compactness) and also the {\it uv}-coverage of the observations. For this reason, the source counts from deep VLBI observations have tended to use the total flux density from lower resolution imaging (e.g.,the VLA; \citealt{Middelberg2013,Herrera_Ruiz2018}) when discussing VLBI-detected radio sources with respect to the radio source population in general. Here, we do use the flux density recovered on VLBI-scales, as we are not primarily interested in testing galaxy formation models or to separate AGN and star-forming galaxies, but instead, our goal is to understand the number of radio sources that could be detected with wide-field VLBI surveys. In Section \ref{discussion}, we will discuss the effect of VLBI resolution bias further and present ways to mitigate it in the future.

\subsection{Euclidean-normalized and differential source counts }

We now calculate the source counts of the radio sources detected during the mJIVE--20 survey. We also compare these with the source counts obtained from studies of compact radio sources selected at higher frequencies. In Table \ref{tab:mjiveNC}, we present the number of sources per flux-density bin, and the correction factor that is determined by taking the effective sky area and completeness of the survey into account. Here, we also present the (corrected) Euclidean-normalized source counts for the mJIVE--20 survey. 

\begin{figure}
 \includegraphics[width=\columnwidth]{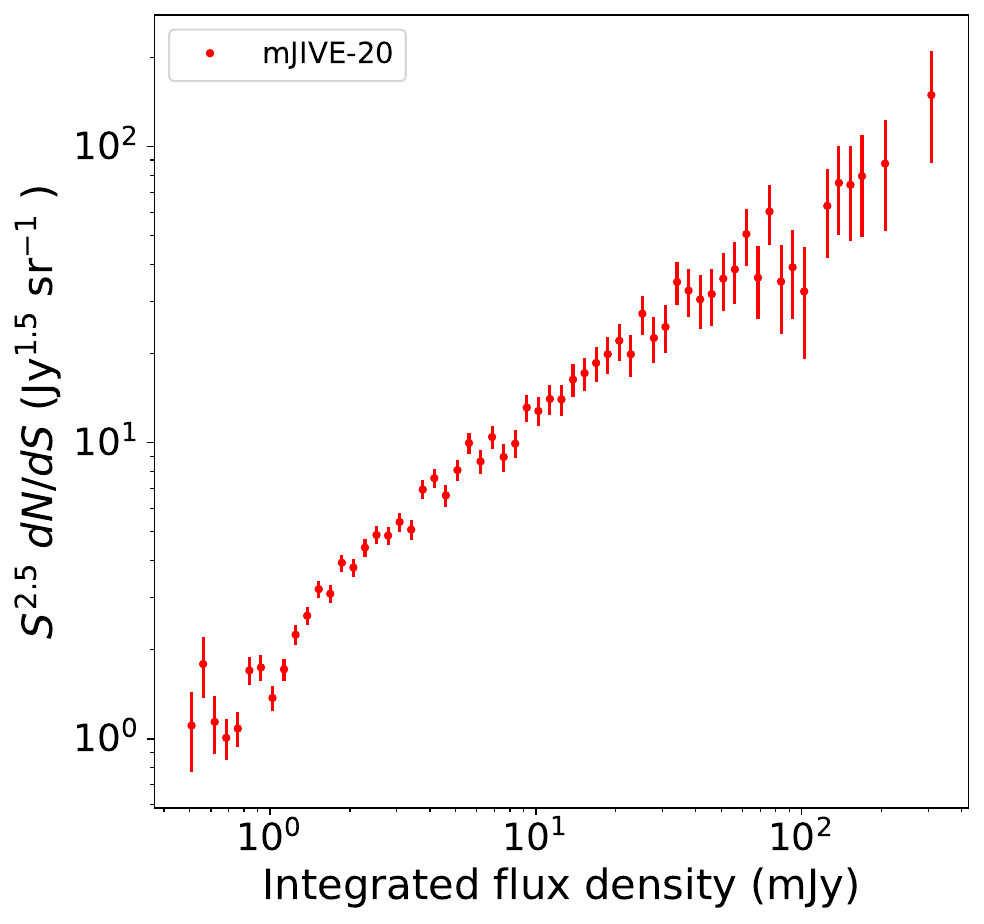}
 \caption{The Euclidean-normalized source counts for VLBI-detected radio sources from the mJIVE--20 survey.}
 \label{fig:NCPL}
\end{figure}

In Fig.~\ref{fig:NCPL}, we present these source counts as a function of flux density, which we see are well approximated by a power-law down to flux densities of around 2 mJy, below which there is evidence of a knee in the distribution. This downturn is likely associated with the change in the parent population of FIRST radio sources described in the previous section. Below 1 mJy, the Euclidean-normalized source counts become flatter, but are also quite noisy. At this stage, it is not clear if this is a systematic or a real effect, similar to the flattening of the radio source counts seen in lower angular resolution observations at this frequency; this is attributed to a change in the radio source population from AGN to star-forming dominated systems \citep{Condon2012,Prandoni2018}. From a power-law fit to the data presented in Fig.~\ref{fig:NCPL}, we find that the Euclidean-normalized source counts can be described by, 
\begin{equation} 
n(S) =  ({2.4 \pm 0.1})  \left(\frac {S_{1.4}}{\rm 1~mJy}\right) ^{({0.695 \pm 0.013})} \quad {\rm Jy^{1.5}~sr^{-1}},
\end{equation}
where $n(S)$ is the differential number of sources (Euclidean-normalized) and $S_{1.4}$ is the flux density at 1.4 GHz.

\begin{figure}
 \includegraphics[width=\columnwidth]{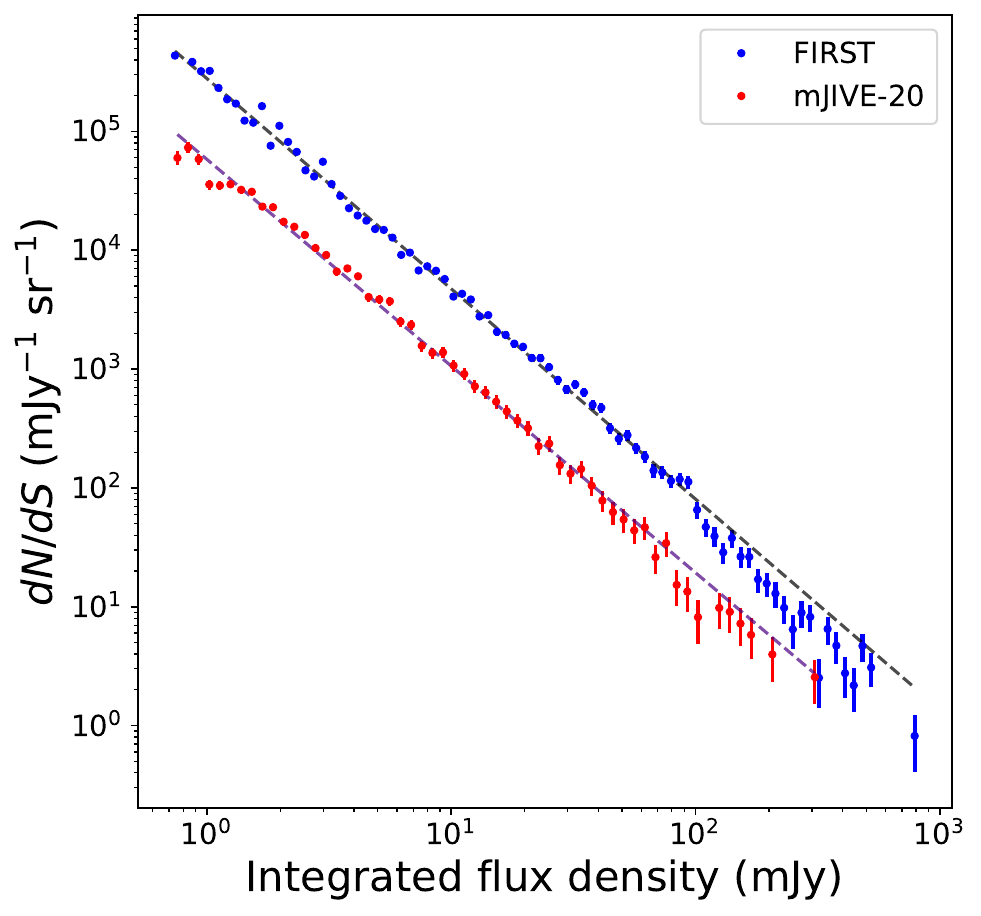}
 \caption{The differential source counts for VLBI-detected radio sources using the mJIVE--20 survey (red), with the FIRST parent population (blue) for comparison. The dashed lines are the best-fit power-law models to the mJIVE--20 survey and FIRST data.}
 %\textbf{Is the integrated flux density here just from one of the surveys or from their respective survey? I'm asking because if it's the latter, then this is a simple translation caused by the average resolving out of the flux.}}
 \label{fig:dNCPL}
\end{figure}

In Fig.~\ref{fig:dNCPL}, we present the differential source counts as a function of flux density for the FIRST parent population and the mJIVE--20 population. From a fit to the data between 1 and 100 mJy, we find that the FIRST parent population can be described by the power-law,

\begin{equation} 
 n(S) =  ({80.6 \pm 5.5}) \left( \frac {S_{1.4}}{\rm 100~mJy}\right)^{-({1.77  \pm 0.02})} \quad {\rm mJy^{-1}~sr^{-1}},
\end{equation}
and that the mJIVE--20 parent population can be described by the power-law,
\begin{equation} 
 n(S) =  ({19.4 \pm 1.4}) \left( \frac {S_{1.4}}{\rm 100~mJy}\right)^{-({1.74 \pm 0.02})} \quad {\rm mJy^{-1}~sr^{-1}}.
\label{eq:nc}
\end{equation}
As in the previous case, $n(S)$ is the differential number of sources and $S_{1.4}$ is the flux density at 1.4 GHz. From these fits, and by inspecting Fig.~\ref{fig:dNCPL}, we see that the power-laws for the parent and mJIVE--20 populations are diverging toward lower flux densities, which would be consistent with the change in the detection fraction seen in Fig.~\ref{fig:mjivedetectionratio}. 
%This is likely due to a change in the source population between the two samples.

Taking the result for the mJIVE--20 survey, we compare these source counts with those for a complete sample of 117 flat-spectrum radio sources selected and observed at 4.85 GHz with the VLA \citep{McKean2007}. Such sources are assumed to be compact (size $< 170$~mas; \citealt{Myers2003}), given their flat-radio spectra (due to the super-position of synchrotron self-absorption from many homogeneous emitting regions) and the higher selection frequency tending to favour core-dominated radio sources. Therefore, flat-spectrum radio sources are prime targets for VLBI observing programmes as they are expected to dominate the compact radio source population. 

We find that the slope of the source counts is steeper [$\eta = -2.06\pm0.01$; where $n(s) = k\,S^{\eta}$], and the normalization is smaller ($k = 6.91\pm0.42$) for flat-spectrum radio sources selected at 4.85 GHz (Jodrell Bank Flat-spectrum radio source sample; JBF; \citealt{McKean2007}), when compared to the results for the VLBI-detected radio sources at 1.4 GHz. However, by assuming a mean spectral index of $\alpha_{1.4}^{4.85} = -0.09$ (where $S_\nu \propto \nu^\alpha$, and $\nu$ is the observing frequency; \citealt{McKean2007}) for the JBF sample, we have calculated the expected number of compact radio sources at 1.4 GHz down to a flux density of 1 mJy. Interestingly, we find that the sky-density of flat-spectrum radio sources selected at 4.85 GHz is very similar to that of the VLBI-detected radio sources from the mJIVE--20 survey, but is a factor $1.22\pm0.12$ higher. Such similar source counts suggests that both samples are drawn from a similar population, even though one is selected based on the radio spectra, and the other on compactness at VLBI-scales. This has implications when choosing an appropriate observing frequency for an all-sky VLBI survey, which we discuss in the next section.

 \begin{table*}
 \centering
  \begin{tabular}{r|r|r|r|r|r}
  \hline 
        $S_l$ & $S_u$ & $\overline{S}$ & Number & Corr.  & Counts\\   
        (mJy)& (mJy)& (mJy)&  & & (sr$^{-1}$~Jy$^{1.5}$) \\ \hline 
0.51 &0.56&0.534&11&11.858 &1.107$\pm$0.334\\
0.57 &0.62&0.595&19&11.858 &1.787$\pm$0.410\\
0.63 &0.68&0.657&20&7.836 &1.140$\pm$0.255\\
0.69 &0.75&0.725&40&5.553 &1.007$\pm$0.159\\
0.76 &0.83&0.799&54&4.416 &1.082$\pm$0.147\\
0.84 &0.92&0.884&86&3.853 &1.700$\pm$0.183\\
0.93 &1.02&0.976&102&3.138 &1.741$\pm$0.172\\
1.03 &1.13&1.080&106&2.460 &1.373$\pm$0.133\\
1.14 &1.24&1.190&131&2.178 &1.715$\pm$0.150\\
1.25 &1.38&1.313&163&1.966 &2.244$\pm$0.176\\
1.39 &1.52&1.454&185&1.798 &2.603$\pm$0.191\\
1.53 &1.68&1.603&224&1.616 &3.198$\pm$0.214\\
1.69 &1.86&1.774&194&1.541 &3.087$\pm$0.222\\
1.87 &2.06&1.961&230&1.427 &3.927$\pm$0.259\\
2.07 &2.27&2.164&203&1.342 &3.785$\pm$0.266\\
2.28 &2.51&2.394&213&1.283 &4.420$\pm$0.303\\
2.52 &2.78&2.652&207&1.231 &4.875$\pm$0.339\\
2.79 &3.07&2.930&190&1.168 &4.852$\pm$0.352\\
3.08 &3.40&3.227&188&1.140 &5.394$\pm$0.393\\
3.41 &3.75&3.582&153&1.118 &5.083$\pm$0.411\\
3.76 &4.15&3.959&186&1.086 &6.940$\pm$0.509\\
4.16 &4.59&4.360&179&1.070 &7.578$\pm$0.566\\
4.61 &5.07&4.854&134&1.057 &6.632$\pm$0.573\\
5.08 &5.61&5.349&143&1.045 &8.073$\pm$0.675\\
5.62 &6.20&5.902&154&1.036 &9.964$\pm$0.803\\
6.21 &6.85&6.527&115&1.033 &8.633$\pm$0.805\\
6.86 &7.57&7.220&119&1.031 &10.441$\pm$0.957\\
7.58 &8.37&7.982&88&1.032 &8.939$\pm$0.953\\
8.38 &9.25&8.791&85&1.030 &9.920$\pm$1.076\\
9.26 &10.23&9.797&94&1.029 &13.135$\pm$1.355\\
10.24 &11.29&10.736&81&1.028 &12.787$\pm$1.421\\
11.33 &12.47&11.896&76&1.028 &14.038$\pm$1.610\\
12.51 &13.80&13.073&66&1.026 &13.986$\pm$1.722\\
13.86 &15.25&14.573&64&1.026 &16.320$\pm$2.040\\
15.28 &16.85&15.982&60&1.026 &17.161$\pm$2.215\\
16.90 &18.65&17.780&55&1.026 &18.571$\pm$2.504\\
18.71 &20.59&19.624&51&1.022 &19.879$\pm$2.784\\
20.68 &22.78&21.701&49&1.023 &22.077$\pm$3.154\\
22.85 &25.15&23.926&38&1.022 &19.873$\pm$3.224\\
25.31 &27.84&26.594&44&1.020 &27.246$\pm$4.108\\
27.98 &30.57&29.148&32&1.020 &22.553$\pm$3.987\\
30.80 &34.00&32.246&30&1.018 &24.583$\pm$4.488\\
34.13 &37.52&35.756&36&1.017 &34.863$\pm$5.811\\
37.66 &41.52&39.615&29&1.016 &32.614$\pm$6.056\\
41.62 &45.72&43.285&24&1.015 &30.452$\pm$6.216\\
45.97 &50.07&48.109&21&1.015 &31.741$\pm$6.926\\
50.93 &55.93&53.434&20&1.012 &35.774$\pm$7.999\\
56.82 &61.93&59.833&18&1.013 &38.466$\pm$9.066\\
62.29 &68.49&65.218&21&1.011 &50.615$\pm$11.045\\
69.47 &74.11&71.762&13&1.008 &36.042$\pm$9.996\\
75.97 &83.62&79.061&19&1.006 &60.303$\pm$13.834\\
84.83 &90.96&87.906&9&1.005 &34.991$\pm$11.664\\
92.70 &100.87&96.787&9&1.007 &39.090$\pm$13.030\\
104.92 &112.86&109.495&6&1.005 &32.393$\pm$13.224\\
125.63 &137.61&132.954&9&1.003 &63.024$\pm$21.008\\
140.59 &151.53&147.273&9&1.003 &75.359$\pm$25.120\\
153.07 &166.88&160.351&8&1.003 &74.221$\pm$26.241\\
169.90 &184.55&179.800&7&1.003 &79.416$\pm$30.017\\
210.57 &225.90&217.562&6&1.004 &87.566$\pm$35.749\\
310.54 &339.24&321.093&6&1.001 &149.152$\pm$60.891\\ \hline
\end{tabular}
  \caption{The Euclidean normalized source counts for the mJIVE--20 survey. The columns from left to right are, the lower ($S_l$) and upper ($S_u$) bounds of a flux density bin, the mean flux density of a bin ($\overline{S}$), the number of sources in each bin ($N$), the correction factor (taking into account the sky area and the completeness; Corr.) and the resulting Euclidean normalized source counts (Counts). Note that 33 sources with flux densities $> 340$~mJy are not included here.}
  \label{tab:mjiveNC}
\end{table*}

\section{Prospects for all-sky VLBI surveys}
\label{discussion}

In this section, we use the differential source counts determined above to estimate the number of radio sources that could be detected from an all-sky survey at mas-scale resolution with VLBI. We focus primarily on what would be achievable with the current VLBA and EVN (using only the smaller dish radio telescopes of the array in a hypothetical survey mode), and briefly discuss the expectations from next generation instruments, like SKA-VLBI. For our calculations, we adopt a survey strategy that maximizes the number of radio sources detected on VLBI-scales.

\subsection{Characteristics of the arrays}

Our ability to survey the sky is dependent on the effective field-of-view of the individual dishes and their sensitivity. This requires some knowledge of the primary beam models, which we assume are consistent with a circular aperture that has an axisymmetric Gaussian illumination pattern, such that the primary beam diameter at the 50 per cent attenuation point is given by,
\begin{equation}
\theta_{\rm FWHM} = 1.22 \lambda / D,
\label{eq:fov}
\end{equation}
where $D$ is the physical diameter of the telescope and $\lambda$ is the observing wavelength. The theoretical sensitivity of an interferometic array is defined from the rms thermal noise fluctuations, given by
\begin{equation}
\sigma_{\rm rms} = \frac{1}{\eta_{\rm cor}}\frac{2 k T_{\rm sys}}{A_{\rm eff}\sqrt{N_p\,N_A(N_A-1) \,\Delta \nu \,\Delta t_{\rm int}}},
\label{eq:sens}
\end{equation}
where $T_{\rm sys}$ is the system temperature (assuming identical antenna systems), $k$ is the Boltzmann constant, $\eta_{\rm cor}$ is the correlator efficiency (assumed to be 0.89 for 2-bit sampling), $A_{\rm eff}$ is the effective collecting area of the filled aperture, $N_p$ is the number of polarisations (assumed to be 2), $N_A$ is the number of antennas in the array, $\Delta\nu$ is the bandwidth in frequency and $\Delta t_{\rm int}$ is the integration time. Typically, the antenna forward gain is defined by the System Equivalent Flux Density (SEFD), which is given by
\begin{equation}
{\rm SEFD} = \frac{2kT_{\rm sys}}{A_{\rm eff}}.
\label{eq:sefd}
\end{equation}
In the cases where the antenna systems are not identical, the SEFD on a given baseline between antennas 1 and 2 is expressed as $\sqrt{{\rm SEFD_1} \times {\rm SEFD_2}}$.  Note that the VLBA and EVN are assumed to operate at central observing frequencies of 1.4 and 1.7 GHz, respectively, given the available frequency coverage of the individual receiver systems.

We see from Equation \ref{eq:fov} that the observable solid angle is proportional to $\lambda^2$, and therefore, surveys of radio sources at longer wavelengths are typically more efficient. This is also because radio sources tend to have a higher flux density toward lower frequencies. However, we see from our comparison of the source counts for the mJIVE--20 and JBF surveys above that the expected number of detectable radio sources on VLBI-scales is likely quite similar at 1.4/1.7 and 5 GHz. Therefore,  either frequency is likely viable. However, the level of radio frequency interference at 5 GHz is currently much less than at 1.4/1.7 GHz, and the usable bandwidth is currently a factor of two to four times better. This, coupled with a slightly better receiver response at 5 GHz, results in around a factor of between 1.5 and 2 improvement in overall sensitivity at 5 GHz. Unfortunately, as this comes at the expense of having to carryout a factor of around 9 to 13 more observations to cover the same area of sky, we focus on 1.4/1.7 GHz as our preferred observing frequency for a potential all-sky survey with VLBI. We note that this choice is based on maximising the number of radio sources detected on VLBI-scales, but other science goals may be better served with the improved angular resolution and frequency coverage afforded at 5 GHz.

\begin{figure*}
    \centering
    \includegraphics[width=\linewidth]{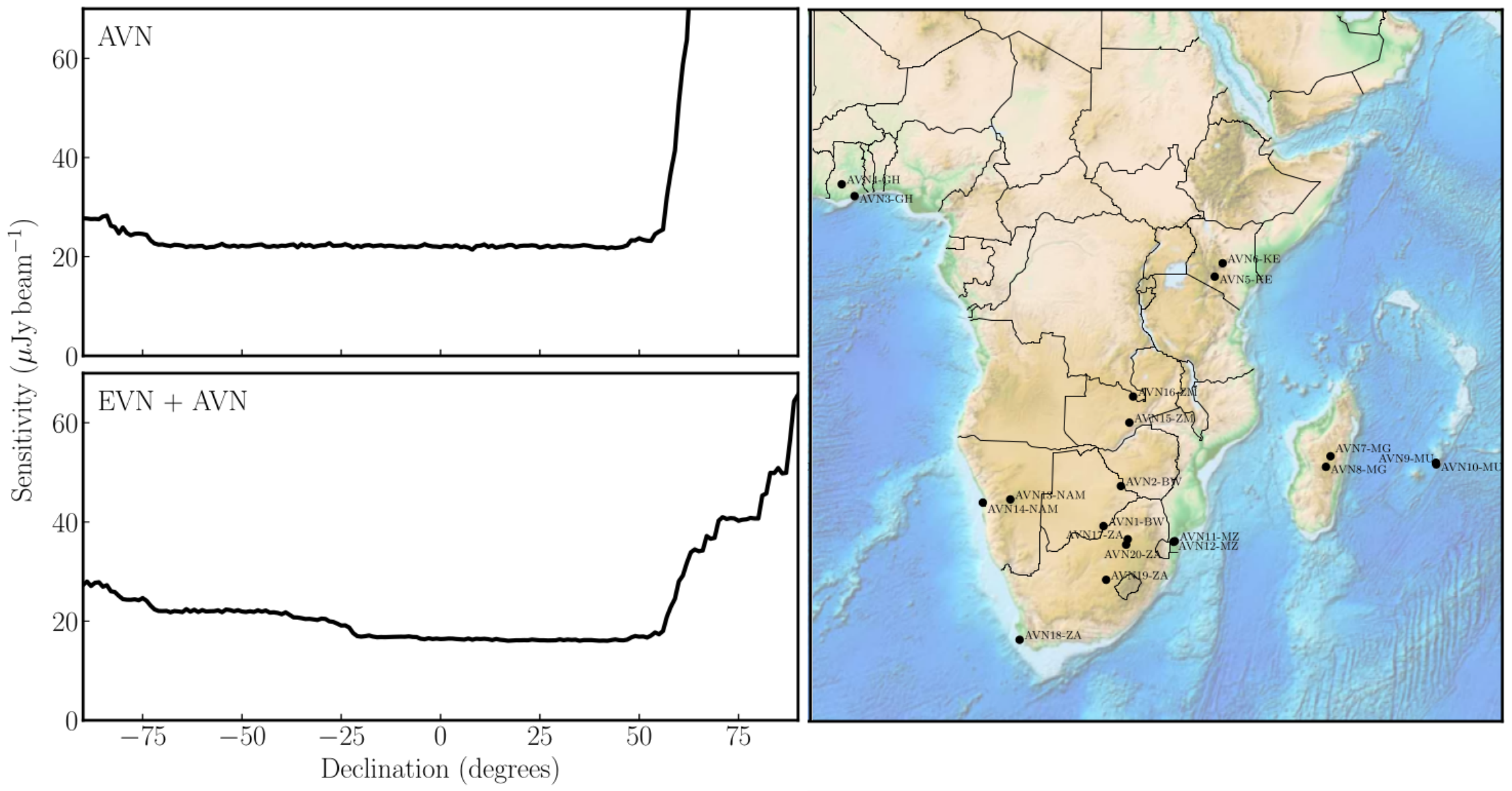}
    \caption{\textit{Left-hand panel:} The rms sensitivity against declination for a theoretical AVN plus EVN array (top) and an AVN-only array (bottom) for a 10 min observation and 512 MHz bandwidth. \textit{Right-hand panel:} The locations of the theoretical 20 element AVN array plotted using \textsc{basemap} \citep{philelson}. The positions of 16 of these telescopes were selected based upon the two largest cities of SKA African partner countries while the remaining are the four largest South African cities.}
    \label{fig:avn_evn}
\end{figure*}

In addition, we also consider a hypothetical VLBI array that includes 20 antennas with the same characteristics as the SKA-MID. We assume that they are positioned in different locations in Botswana, Ghana, Kenya, Madagascar, Mauritius, Mozambique, Namibia and Zambia (the partner African VLBI Network countries) with two dishs per country, and a further four antennas spread across South Africa. The exact locations of these antennas can be seen in the right-hand panel of Figure~\ref{fig:avn_evn} Also, we assume that 512 MHz of frequency bandwidth will be usable within the SKA Band 2, operating between 0.95 and 1.76 GHz, due to radio frequency interference.

In Table \ref{tab:arrays}, we summarise the properties of the three arrays that we consider here.

\begin{table*}
\centering
 \begin{tabular}{l l p{6.5cm} l l l l l} \hline
    Array           &   $N_A$   & Antennas                                                  & Diameter  & Freq. & $\Delta\nu$ & FoV         & SEFD \\
                    &           &                                                           & (m)       & (GHz) & (MHz)       & (deg$^2$)   & (Jy) \\ \hline
    VLBA            &  10       & Sc, Hn, Nl, Fd, La, Kp, Pt, Ov, Br, Mk                    & 25        & 1.4   & 256         & 0.262       & 365 \\
    EVN/e-MERLIN    &  18       & Mc, On, Tr, W1, Nt, Sh, Ur, Hh, Sv, Zc,\\
    & & Bd, Ir, Jb2, Cm, Da, De, Kn, Pi   & 25, 32    & 1.7   & 128         & 0.117       & 485 \\  
    SKA-VLBI        &  20   & AVN1-BW, AVN2-BW, AVN3-GH, AVN4-GH, AVN5-KE, AVN6-KE, AVN7-MG, AVN8-MG, AVN9-MU, AVN10-MU, AVN11-MZ, AVN12-MZ, AVN13-NAM, AVN14-NAM, AVN15-ZM, AVN16-ZM, AVN17-ZA, AVN18-ZA, AVN19-ZA, AVN20-ZA & 15  & 1.4 & 512 & 0.728 &  235 \\
    \hline
    \end{tabular}
    \caption{The properties of the two arrays considered here for an all-sky survey with VLBI. The antenna names are given using their standard abbreviation. The portion of EVN/e-MERLIN listed here has a combination of $10\times 25$- and $8\times32$-m telescopes, but for our field-of-view (FoV) calculations, we use a diameter of 32 m. The SEFD is the average for the given array. We also include a hypothetical VLBI array that includes 20 antennas with the characteristics of the SKA-MID dish design, which provides mas-scale angular resolution (excluding the SKA-MID core).}
    \label{tab:arrays}
\end{table*}

\subsection{Prospects for in-beam calibration}

Our ability to calibrate VLBI observations is currently limited by the density of objects that are suitable for correcting the complex antenna gains as a function of time. Also, given the large changes in phase, due to the long baselines involved with VLBI observations, calibrators are often required to be very close to the target field ($< 2$~deg separation). There are currently 17\,432 compact radio sources that have been identified as calibrators for VLBI experiments (1 in 2.37 deg$^{2}$ over the whole sky; Radio Fundamental Catalog; \citealt{Petrov2021}). However, astrometric accuracy scales with the calibrator distance \citep{Pradel2006}, so additional calibrators will improve the overall astrometric and imaging quality. It is this rationale that made in-beam calibration ($<0.4$~deg calibrator-to-target separation) a key component of the mJIVE--20 survey. 

With the development of wide-bandwidth observations, radio sources at lower flux densities can now be used as calibrators, when the amplitude and phase of the visibilities are well calibrated as a function of frequency. For example, the recommended phase referencing time at 1 to 8 GHz in good weather is 300 s. For the VLBA and EVN arrays presented in Table~\ref{tab:arrays}, this is equivalent to a baseline sensitivity of $\sigma_{\rm rms} = 0.74$ and 0.98~mJy~beam$^{-1}$ at 1.4 and 1.7 GHz, respectively. Therefore, compact radio sources with flux-densities $>7.4$ and $>9.8$~mJy (for $10\sigma_{\rm baseline}$) can be regarded as good phase reference sources for the VLBA and EVN, respectively. For our hypothetical SKA-VLBI array, the baseline sensitivity is $\sigma_{\rm rms} = 0.39$~mJy~beam$^{-1}$, which corresponds to a minimum calibrator flux density of $>3.9$~mJy (for $10\sigma_{\rm baseline}$).

Using the differential source counts of VLBI-detected radio sources given by equation \ref{eq:nc}, we calculate the expected sky density of radio sources needed for phase referencing from pointed observations, and determine the fraction that can be used for in-beam calibration. Note that the latter requires increasing the lower limit on the flux density by a factor of 2, due to the attenuation of the primary beam of the antennas. For the VLBA, we find that the sky density of radio sources with flux densities $>7.4$ (pointed) and $>14.8$~mJy (in beam) is 5.5 and 3.3~deg$^{-2}$, respectively. This corresponds to an in-beam calibrator density of 0.87~sources~beam$^{-1}$. In the case of the EVN, we expect that the sky density of radio sources with flux densities $>9.8$ (pointed) and $>19.6$~mJy (in beam) is 4.5 and 2.6~deg$^{-2}$, respectively, with an in-beam density of 0.30~sources~beam$^{-1}$ (for the 32-m antennas). Finally, for our hypothetical SKA-VLBI array, the sky density of radio sources with flux densities $>3.9$ (pointed) and $>7.8$~mJy (in beam) is 8.8 and 5.3~deg$^{-2}$, respectively. Therefore, the number of radio sources that can be used for in-beam calibration with the SKA-VLBI array would be 3.9~sources~beam$^{-1}$.

Overall, we find that 1 in 3 pointings with the EVN will likely have an in-beam calibrator, whereas this is likely the case for almost all pointings with the VLBA \citep{Ding2023, Deller2011B}. In the case of the SKA-VLBI array, we predict that there will be multiple in-beam calibrators available, which can be important for extremely precise astrometric applications of VLBI (e.g., \citealt{Dodson2018}).

\subsection{Expected number of detected sources}

We now calculate the number of radio sources that can be found from an all-sky survey with VLBI. For this, we integrate equation \ref{eq:nc} to an appropriate flux-density limit. We see from Fig.~\ref{fig:NCPL} that our differential source counts of VLBI-detected radio sources is robust to around 1 mJy, below which the source counts become much more noisy. In principle, we could extrapolate our calculations below 1 mJy, but given the expected change in the radio source population at these flux densities, this would likely result in an over-estimate of the total number of radio sources that we would detect. Therefore, we use a 1.4/1.7 GHz flux-density limit of 1 mJy for our calculations.

We also assume that, given the short amount of time used per observation, any survey would have similar noise properties to the mJIVE--20 survey, and that the same detection threshold would likely be needed ($6.75\sigma$); this would result in a similar completeness of 97 per cent, as we determined from Fig.~\ref{fig:blobcat_mjivesetting}, which we fold into our calculations. We also assume that the individual pointings are separated by a beam width, in a similar configuration to what was presented in Fig.~\ref{fig:area}. In a future work, we will simulate the effect of different pointing strategies and noise realizations. However, for our current pointing strategy, we see that the absolute value of the noise changes across the field. Therefore, to make a $6.75\sigma$ detection at a flux density (point-source) limit of 1 mJy, means having an average thermal noise of 150~$\mu$Jy~beam$^{-1}$. This corresponds to a thermal noise at the pointing centre (where the primary beam response is maximum) of 130~$\mu$Jy~beam$^{-1}$ ($1\sigma_{\rm rms}$). From the properties of the three arrays given in Table \ref{tab:arrays}, and using equations \ref{eq:sens} and \ref{eq:sefd}, we find that the on-source integration time needed to reach this sensitivity
would be 215, 225 and 15 s for the VLBA, EVN and a hypothetical SKA-VLBI array, respectively. This highlights the improvement in sensitivity that is potentially available with SKA-VLBI, given the advances in antenna and receiver design for this next generation instrument. We note that the short integration times needed to reach the required noise levels would allow for flexible scheduling of such observations. Therefore, it would also be straightforward to define optimum scheduling blocks when the target fields are visible to all of the available telescopes. This is more challenging to achieve for long-track observations that are typically used for deep-field science cases.

We also see from equation \ref{eq:nc} that the slope of the differential number counts of VLBI-detected radio sources is quite shallow ($\eta = -1.74\pm0.02$). Therefore, maximizing the number of detected radio sources is best achieved through observing the largest possible  sky area, which for our calculations we assume is $3\pi$~sr (equivalent to about 31\,000~deg$^2$). Given a flux-limited survey to 1 mJy, and a completeness of 97 per cent at this limit, such a survey could potentially detect $(7.2\pm0.9)\times10^{5}$~radio sources on VLBI-scales, a factor of about 30 times more than is currently known.

Finally, we estimate the total time such a survey would take with the current VLBA and EVN (using only the 32-m antennas), and a next generation instrument, like SKA-VLBI. From Table~\ref{tab:arrays}, we see the effective field-of-view of each array. From this we estimate that to survey $3\pi$~sr would take $1.18\times10^5$, $2.64\times10^5$ and $1.18\times10^4$ pointings for the VLBA, EVN and a hypothetical SKA-VLBI array, respectively. Given the integration time to reach the required noise level, this equates to about 7000, 16500 and 180 h of on-source observing time for the VLBA, EVN and SKA-VLBI, respectively. Note that about 2 in 3 of the EVN pointings will also require additional phase referencing calibration (see above), which will add up to around another 30 per cent to the total time needed to complete such a survey with that array. For this reason, such surveys with the EVN are likely prohibitive, whereas with the VLBA and SKA-VLBI they would be very much feasible.

\section{Conclusions}
\label{conclusions}

We have analyzed the final catalogue of the mJIVE--20 survey, which observed 24\,903 radio sources in the FIRST survey using the VLBA at 1.4 GHz to an rms noise level of about 150~$\mu$Jy~beam$^{-1}$, detecting 4\,965 radio sources on VLBI-scales. Through comparing the number of detections and non-detections in the mJIVE--20 catalogue, we found that the detection fraction is a strong function of the peak surface brightness (compactness and radio source size) of the objects in the FIRST survey; the detection fraction is over 50 per cent at a peak surface brightness at 80~mJy~beam$^{-1}$ and falls steadily to about 31 per cent at 5 mJy~beam$^{-1}$. Below this, the detection fraction falls sharply to 8 per cent at the detection threshold of both surveys. This is likely due in part to a change in the composition of the radio source population, from AGN to star-formation dominated objects, but is also due to the VLBI observations not being sensitive enough to detect low-level compact emission toward lower flux densities. We found an overall detection fraction of $19.9\pm2.9$~per cent. Given the limited {\it uv}-coverage of the mJIVE--20 survey observations, we found that those radio sources that are detected tend to be unresolved (median VLBI compactness 0.9~beam$^{-1}$), with a median size of 7.7~mas (about half a beam-size). Finally, we found that 20 to 35 per cent of the radio sources detected were resolved, with sizes $>16$~mas.

From an analysis of the VLBI-detected source counts, we see hints of a similar behaviour in the Euclidean-normalized distribution that has been reported on arcsec-scales, that is, a downturn around a flux density of 2 mJy and then a potential flattening below 1 mJy. This might arise from the considerable challenge in accurately determining the size and, by extension, the integrated flux density of sources with low signal-to-noise ratios. During the deconvolution process, if the calculated size is smaller than the actual beam size, the resulting integrated flux could be significantly lower than the peak flux density. As a result, these measurements are accompanied by substantial error bars. We also determine the differential number counts for VLBI-detected radio sources, finding that the total number of objects is similar to those of flat-spectrum radio sources selected at higher frequencies, suggesting that they come from a similar population.

From our analysis of the differential source counts, we found that the sky density of suitable phase reference sources is of order 2.6 to 3.3~deg$^{-2}$. This should be sufficient for in-beam phase referencing in the case of around 30 and 90 per cent of the observations carried out with the EVN and VLBA, respectively. However, we found that for a VLBI facility that includes antennas with a similar specification to the SKA-MID design, the expected sky density of phase reference sources is about 5.3~deg$^{-2}$, which equates to multiple in-beam calibration sources for each observation.

Finally, we investigated the number of sources that could be found from all-sky surveys carried out with the VLBA and EVN. From this analysis, we found that a factor 30 more VLBI-detected radio sources could be identified with around 7000 h of observations with the VLBA. However, in the case of the EVN, such surveys would be rather expensive, given the smaller field of view of the antennas in that array. For a hypothetical SKA-VLBI array, such a survey would take a fraction of the time, and could be completed with just 180 h of observations. Our analysis is currently limited by our knowledge of the radio source counts below 1 mJy. More focused surveys of the radio sky with VLBI, down to a limiting sensitivity of around 100~$\mu$Jy~beam$^{-1}$ ($6\sigma$ detection threshold) would be extremely informative in testing models for radio source populations and making robust predictions for the expectations with SKA-VLBI.

\section*{Acknowledgements}
This paper is based on research developed in the  DSSC Doctoral Training Programme, co-funded through a Marie Skłodowska-Curie COFUND (DSSC 754315). JPM acknowledges support from the Netherlands Organization for Scientific Research (NWO) (Project No. 629.001.023) and the Chinese Academy of Sciences (CAS) (Project No. 114A11KYSB20170054). The National Radio Astronomy Observatory is a facility of the National Science Foundation operated under cooperative agreement by Associated Universities, Inc.

\section*{Data Availability}
The data used in this research is publicly available in the mJIVE--20 survey and FIRST databases.

%%%%%%%%%%%%%%%%%%%% REFERENCES %%%%%%%%%%%%%%%%%%

% The best way to enter references is to use BibTeX:

%\bibliographystyle{mnras}
%\bibliography{example} % if your bibtex file is called example.bib

% Alternatively you could enter them by hand, like this:
\bibliographystyle{mnras}
\bibliography{example} % if your bibtex file is called example.bib

%%%%%%%%%%%%%%%%%%%%%%%%%%%%%%%%%%%%%%%%%%%%%%%%%%

%%%%%%%%%%%%%%%%% APPENDICES %%%%%%%%%%%%%%%%%%%%%

\appendix
\section{Hypothetical AVN locations}

\begin{table*}
\centering
\caption{The geocentric coordinates of a hypothetical African VLBI Network using 20 SKA-MID dishes.}
\begin{tabular}{ccccc}
\hline
\multicolumn{3}{c}{Geocentric coordinates} & & \\
$x$ & $y$ & $z$ & Station code & Country \\
(m) & (m) & (m) & & \\
\hline
$5217030.16853176$     & $2534221.5108732083$ & $-2644582.0758971698$ & AVN1-BW & Botswana \\
$5277247.922210887$ & $2748624.8644503574$ & $-2289328.600036085$ & AVN2-BW & Botswana \\
$6347825.285561881$ & $-20717.88800793007$ & $618657.4567346348$ & AVN3-GH & Ghana \\
$6332560.529802635$ & $-178731.63840587594$ & $737359.7953479206$ & AVN4-GH & Ghana \\
$5104748.927307364$ & $3821230.688393553$ & $-142231.03817239657$ & AVN5-KE & Kenya \\
$5049932.002818037$ & $3895996.353310751$ & $5528.713103446548$ & AVN6-KE & Kenya \\
$4075522.7134878747$ & $4450989.035238711$ & $-2056922.0894522057$ & AVN7-MG & Madagascar \\
$4090030.2866659025$ & $4391134.402974519$ & $-2153824.1004977995$ & AVN8-MG & Madagascar \\
$3218155.6429566364$ & $5051492.497668295$ & $-2185029.212682493$ & AVN9-MU & Mauritius \\
$3213616.693183972$ & $5047594.365627962$ & $-2200557.8274866333$ & AVN10-MU & Mauritius \\
$4837312.3501758$ & $3090783.6593368473$ & $-2770624.9059733003$ & AVN11-MZ & Mozambique \\
$4841119.498520394$ & $3080179.155624241$ & $-2775745.617128389$ & AVN12-MZ & Mozambique \\
$5632553.87753738$ & $1731035.1721495497$ & $-2432816.736449252$ & AVN13-NAM & Namibia \\
$5688944.726890755$ & $1471992.094229581$ & $-2471560.196273584$ & AVN14-NAM & Namibia \\
$5415867.714182049$ & $2914108.406276784$ & $-1684593.4154433303$ & AVN15-ZM & Zambia \\
$5456235.737166995$ & $2979000.8406833992$ & $-1421987.8923449807$ & AVN16-ZM & Zambia \\
$5066809.931828817$ & $2715444.8969223998$ & $-2753782.8449464357$ & AVN17-ZA & South Africa \\
$5026325.59405273$ & $1674472.17663092$ & $-3539546.651210291$ & AVN18-ZA & South Africa \\
$5003701.377073657$ & $2463938.124620917$ & $-3083590.4930300973$ & AVN19-ZA & South Africa \\
$5054121.780624786$ & $2692033.063658318$ & $-2799397.734311496$ & AVN20-ZA & South Africa \\
\hline
\end{tabular}

\end{table*}

%%%%%%%%%%%%%%%%%%%%%%%%%%%%%%%%%%%%%%%%%%%%%%%%%%

% Don't change these lines
\bsp	% typesetting comment
\label{lastpage}
\end{document}